\documentclass[12pt,preprint]{aastex}

\shorttitle{Swift UVOT Observations of GRB050319}
\shortauthors{K. O. Mason et al.}
\begin{document}
\title{Prompt optical observations of GRB050319 with the Swift UVOT}
\author{K. O. Mason$^1$, A. J. Blustin$^1$,  P. Boyd$^2$, S. T. Holland$^2$, M. J. Page$^1$, P. Roming$^3$, M. Still$^2$, 
B. Zhang$^4$, A. Breeveld$^1$, M. De Pasquale$^1$, N. Gehrels$^2$, C. Gronwall$^3$, 
S. Hunsberger$^3$, M. Ivanushkina$^3$, W. Landsman$^2$, K. McGowan$^1$, J. Nousek$^3$, 
T. Poole$^1$, J. Rhoads$^5$, S. Rosen$^1$, \& P. Schady$^1$  
} 
\affil{$^1$Mullard Space Science Laboratory, Department of Space and Climate
Physics, \\ University College London, Holmbury St Mary, Dorking, Surrey,
RH5 6NT, UK.\\
$^2$ NASA/Goddard Space Flight Center, Greenbelt, MD 20771, USA.\\
$^3$ Department of Astronomy and Astrophysics, Pennsylvania State University, 525 Davey Laboratory, University Park, PA 16802, USA.\\
$^4$ Department of Physics, University of Nevada, Las Vegas, Nv 89154, USA.\\
$^5$ Space Telescope Science Institute,3700 San Martin Drive, Baltimore, Md 21218 \\
} 
\email{kom@mssl.ucl.ac.uk}
\received{2005 June 16}
\begin{abstract}
The UVOT telescope on the Swift observatory has detected optical afterglow emission from GRB 050319. The flux 
declines with a power law slope of $\alpha = -0.57$ between the start of observations some 230 seconds after the
burst onset (90s after the burst trigger) until it faded below the sensitivity threshold of the instrument after
$\sim 5 \times 10^4$s. There is no evidence for the rapidly declining component in the early light curve that is seen
at the same time in the X-ray band. The afterglow is not detected in UVOT shortward of the B-band, suggesting a 
redshift of about 3.5. The optical V-band emission lies on the extension of the X-ray spectrum, with an optical
to X-ray slope of $\beta = -0.8$. The relatively flat decay rate of the burst suggests that the central engine
continues to inject energy into the fireball for as long as a few $\times 10^4$s after the burst.
\end{abstract}


\clearpage

\section{Introduction}
The Swift observatory (Gehrels et al. 2004) is designed to localise Gamma-ray bursts on the sky rapidly 
and then bring its X-ray and ultraviolet/optical  
telescopes to bear on that location within about one minute to provide panchromatic observations 
of the bursts 
and their afterglows. Initial information on the burst location and properties is transmitted 
to the ground in near real time for distribution to follow-up observers. In this letter we describe
Swift observations of the burst GRB050319, concentrating on the data from the Ultraviolet and Optical
Telescope. 

\section{Observations and Analysis}
The Swift Burst Alert Telescope (BAT; Barthelmy et al. 2005) triggered on the Gamma-ray Burst 
GRB050319 at 09:31:18.44 UT (Krimm et al. 2005a).   
The burst lightcurve was initially reported as having a single peak with duration  
T90 = 10 {$\pm$} 2 seconds (Krimm et al. 2005b). However subsequent analysis has shown that
this was a multi-peaked burst that actually began about 138s before the Swift trigger (Figure~1). 
The burst exhibits several peaks during an initial period of about 80 seconds,
followed by a similar length interval during which the BAT flux is indistinguishable from background, and then
the final peak that generated the BAT trigger. The BAT did not trigger on the earlier activity because the
spacecraft was slewing; triggers are disabled during slews. The slew ended about 50s before the
final peak. The energy index
of the 1-s peak spectrum (starting at the trigger time, T, +0.36 sec.) is $\beta = -1.1$ {$\pm$} 0.3
(90\% confidence; $\beta$ defined such that $f_{\nu}\propto 
\nu^{\beta}$).  The time-averaged spectrum of the final peak yields an energy index of
--1.2 {$\pm$} 0.2 (Krimm et al. 2005b). 
 

\begin{figure}
\epsscale{0.8}
\plotone{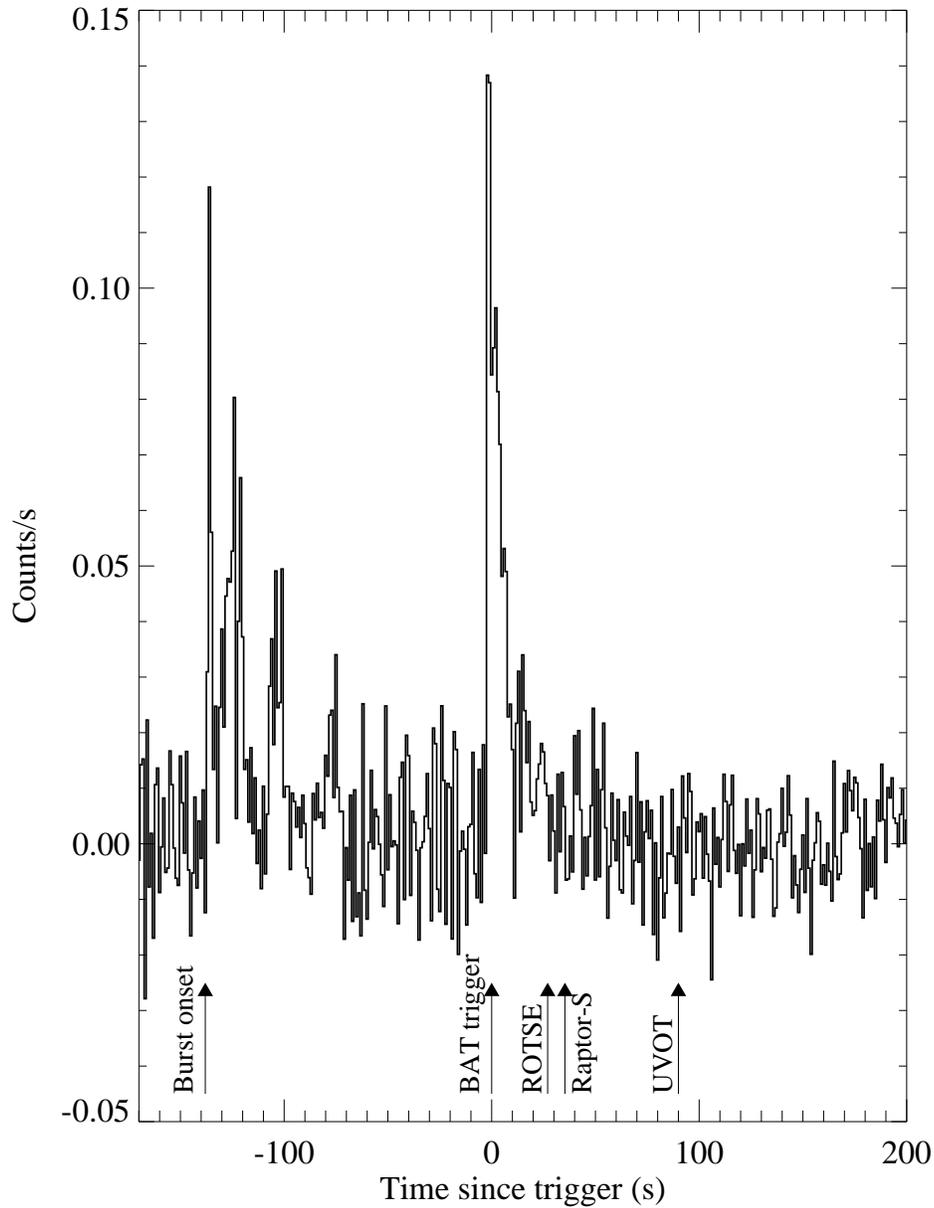}
\caption{The BAT mask-weighted light curve of GRB050319 in the 15-350 keV band at 1 second 
time resolution. The burst began 138 seconds before the BAT trigger, which occured on final 
peak of the burst. The spacecraft was slewing at the time of the previous peaks, 
and BAT triggers are disabled during slews. The slew ended about 50s before the trigger. The start of
coverage from the ROTSE (Quimby et al. 2005) and Raptor-S (Wozniak et al. 2005) ground-based robotic telescopes
and the Swift UVOT are marked. 
}
\end{figure}

\begin{figure}
\epsscale{1.0}
\plotone{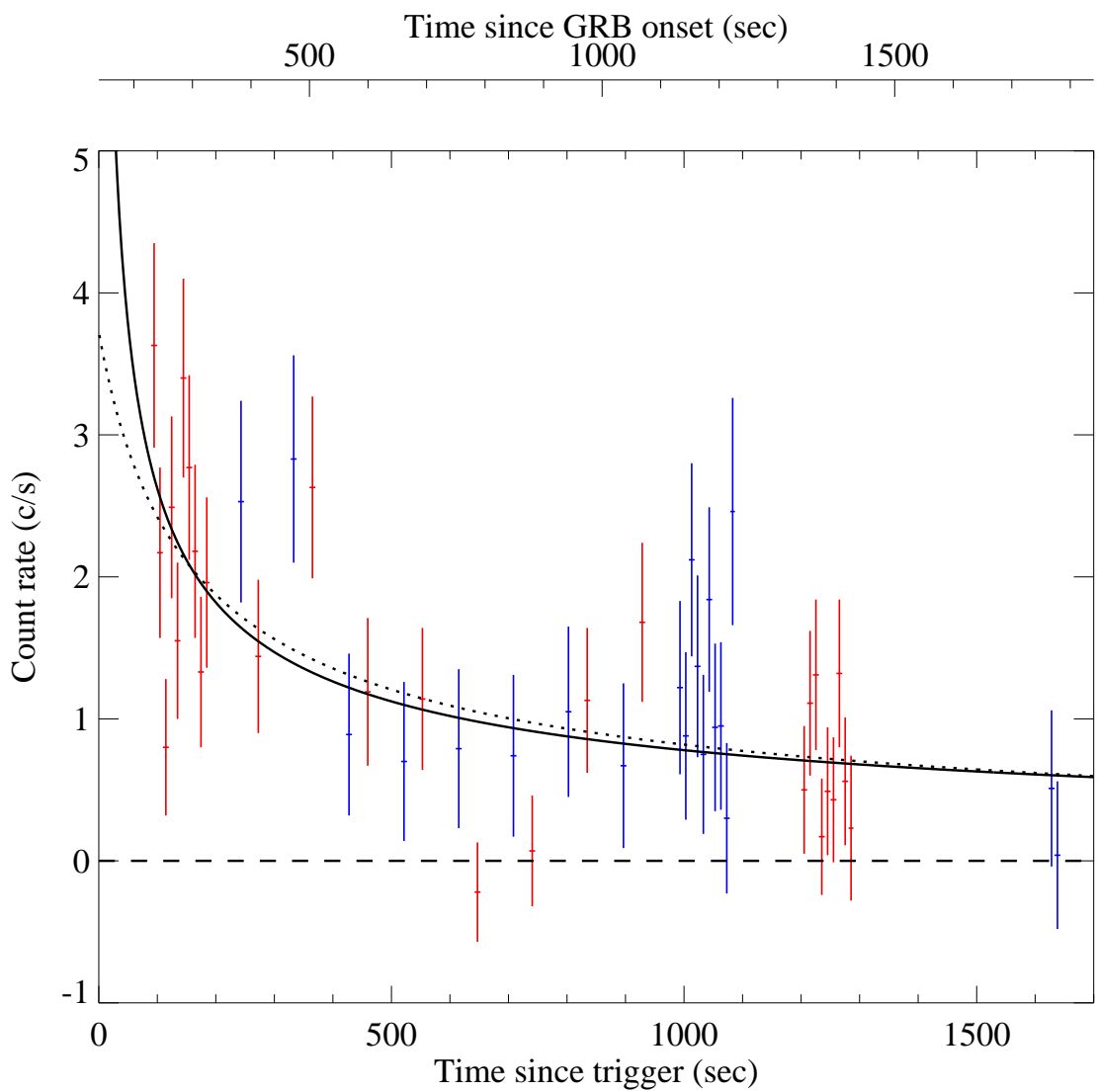}
\caption{UVOT V-band (red) and B-band (blue) count rate data in the interval up to 1700s after the burst trigger.
Since as explained in the text, the burst began about 138s before the trigger, we indicate time since
burst onset on the upper axis.
The B-band data have been multiplied by a factor 0.8 to normalise them to the V-band (see text). 
The solid curve is the best fit power law decay curve, which has  $\alpha = -0.57$. The dashed curve is a fit in
which the time of origin of the power law, $T_0$, is allowed to be a free parameter (see text). 
}
\end{figure}

\begin{figure}
\epsscale{0.8}
\plotone{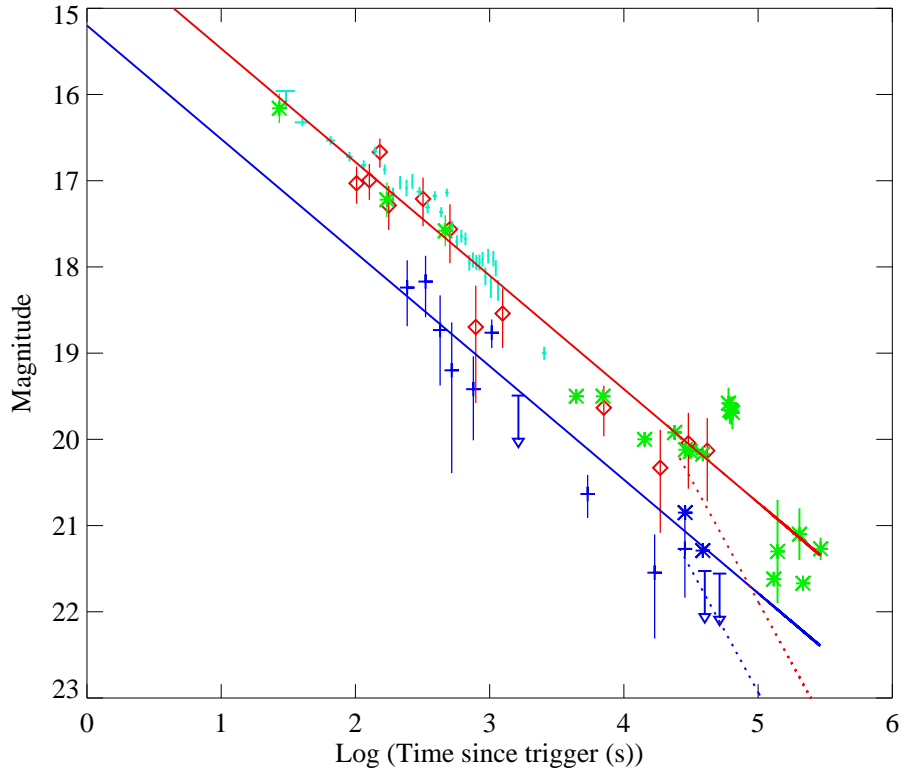}
\caption{Measurements of the GRB050319 afterglow decay plotted as magnitudes versus the log of the time interval
since the burst trigger. UVOT V-band (red diamonds) and B-band (blue crosses) magnitudes are compared against
groundbased data reported in the literature (see text). The ground-based data are predominantly R-band, or equivalent R-band,
measurements (green stars), but two groundbased B-band measurements (blue stars) are also shown. We also show the
equivalent R-band measurements of the early afterglow reported by Wo\'zniak et al. (2005) (cyan).  A power law
decay curve with a slope $\alpha = -0.57$ is compared to the V and B data. The dashed lines show a break to an 
$\alpha = -1.14$ slope as suggested by analysis of the X-ray data on this burst (Cusumano et al. 2005).
}
\end{figure}

\begin{figure}
\plotone{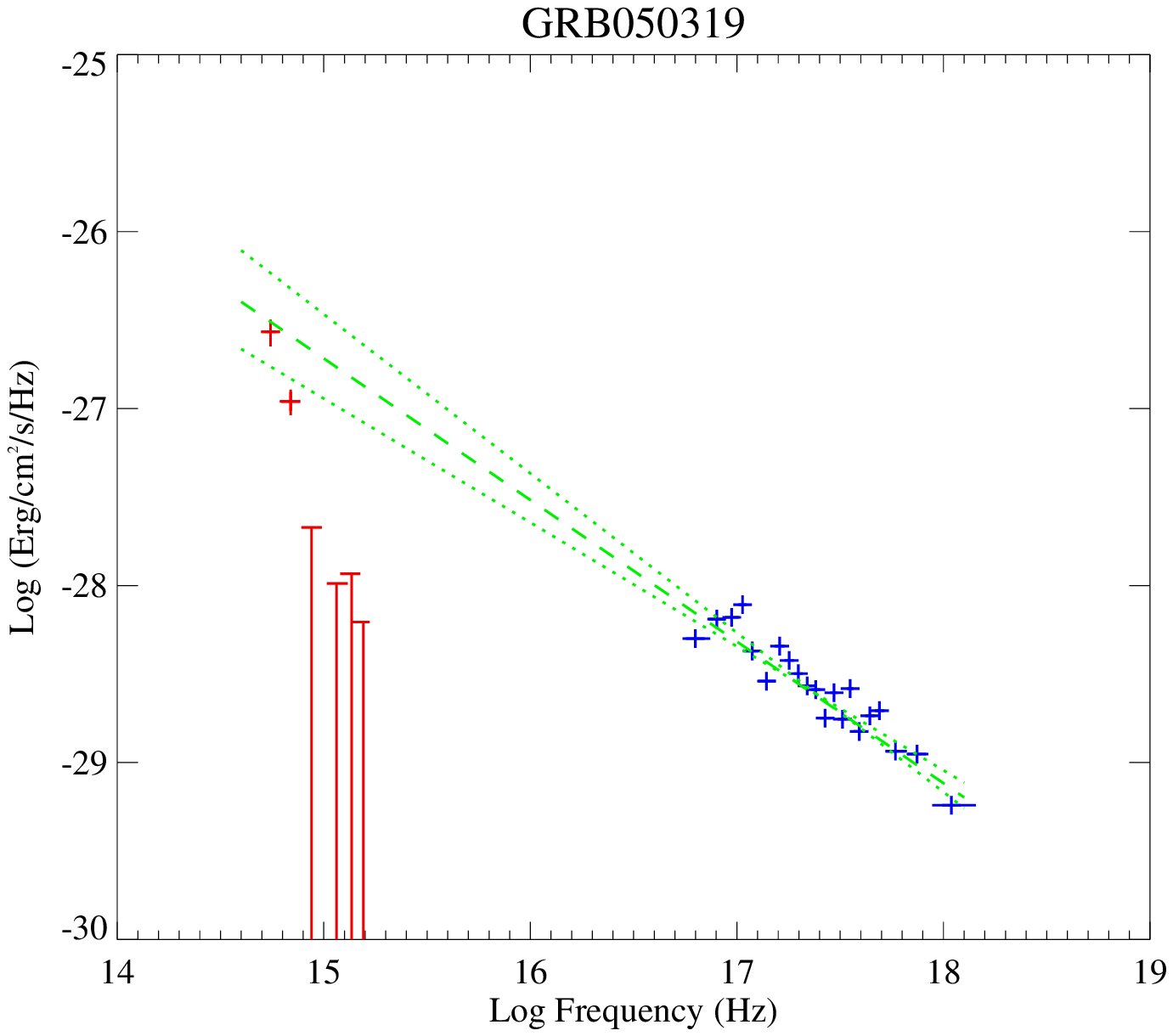}
\caption{The multiwavelength spectrum of GRB050319 plotted as F$\nu$ vs log frequency in the observer frame, 
and averaged over the interval between 240s and 930s after the burst trigger. UVOT points are shown in red, XRT data in blue.
The best power law fit to the XRT data is illustrated as the green dashed line. The 1$\sigma$ bounds on the slope are
indicated by the dotted lines.
The XRT flux points have been corrected for
Galactic absorption equivalent to N$_{\rm H} = 1.17 \times 10^{20}$ cm$^{-2}$.
}
\end{figure}

The Swift spacecraft executed a slew immediately to bring the narrow field X-ray Telescope (XRT; Burrows et al. 2005) and
Ultraviolet/Optical Telescope (UVOT; Roming et al. 2005) to bear on GRB050319, and observations with these started about
90 seconds after the burst trigger (230s after the burst onset). In this paper we concentrate on the analysis of the UVOT 
data and compare with the results from the XRT observations reported by Cusumano et al. (2005). 
We also consider data from a number of ground-based observatories that responded to 
the Swift trigger (Yoshioka et al. (2005), Quimby et al (2005), Kiziloglu et al. (2005), Misra et al. (2005), 
Sharapov et al. (2005a,b), Greco et al. (2005) and Wo\'zniak et al (2005)). The earliest observations were made 27s 
after the Swift trigger (Quimby et al. 2005). 

\subsection{Optical decay}

The Swift UVOT executed a standard series of exposures, the first of which was a `finding chart' 
exposure of duration 100s in the V filter that began 90s after the BAT trigger. 
Thereafter, the instrument cycled through each of six color 
filters, V, B and U 
together with filters defining three ultraviolet passbands, UVW1, 
UVM2 \& UVW2 (see Table~2 for central wavelengths). The exposure duration per filter was initially 10 seconds, subsequently increasing to 100s and then 900s at 
predetermined times after the trigger. The initial data (up to 1700s after the trigger) 
were taken in `event mode' in which the time and detector position of each 
individual photon is recorded. For the later exposures, the instrument was switched to `image mode', in which the image is accumulated
onboard to conserve telemetry volume, discarding the photon timing information within an exposure. The intrinsic spatial 
pixel size of the detector is approximately 0.5 arc seconds on the sky, but the image mode data were taken with the data binned 2x2 
to give 1 arc second pixels.

\begin{deluxetable}{cccll}
\tablewidth{0pc}
\tablecaption{UVOT Time Resolved Photometry}
\tablehead{
\colhead{Mid Time$^1$}  & \colhead{Exposure}  & \colhead{Magnitude} & 
\colhead{Error}   
 \\
\colhead{(s)}  & \colhead{Time}  & \colhead{} & 
\colhead{(Mag)}  & \colhead{} 
 \\
\colhead{}  & \colhead{(s)}  & \colhead{} & 
\colhead{}  & \colhead{}
 \\

}
\startdata
{\it V Band}\\
102	& 25	&	17.03 &	0.22\\	   
127	& 25	&	16.99 &	0.21\\	   
152	& 25	&	16.67 &	0.17\\	   
177	& 25	&	17.28 & 0.26\\	   
319	& 20	&	17.21 & 0.28\\	   
506	& 20	&	17.56 &	0.34\\	   
788	& 40	&	18.70 &	0.69\\	   
1249	& 100	&	18.54 & 0.35\\	       
7088	& 700	& 19.65	& 0.29\\	   
18664 &	689	& 20.35	& 0.57	\\   
30217 &	726	& 20.07	 & 0.42	  \\ 
41791 &	718	& 20.09	& 0.43	 \\
\\
{\it B Band}\\
243	   &     10 &	18.24 & 0.38 \\	   
333	&        10	&	18.17	 & 0.36	\\   
428	&        10	&	18.73	& 0.52	\\   
522	&        10	&	19.19 &	0.86	\\   
756	&        40	&	19.42 & 0.48\\	   
1036	&       100	&	18.76 & 	0.17	\\   
1637	&        26	&	$>$19.5 \\	 
5370	&       900	&	20.63 & 0.25\\	   
16951   &	900	&	21.54 & 0.60\\	   
28491   &	897	&	21.27 & 0.47\\	   
40064   &	900	&	$>$21.5 \\	   
51638   &	900	&	$>$21.5 \\
\enddata
\tablecomments{$^1$ Time since 2005 Mar 19 9hr 31m 18.44s
}
\end{deluxetable}

\begin{deluxetable}{lcclcl}
\tablewidth{0pc}
\tablecaption{UVOT Spectral Photometry: T+(240s-930s)}
\tablehead{
\colhead{Filter}  & \colhead{Effective}  & \colhead{Integrated} & 
\colhead{Counts/s}  & \colhead{Zeropoint} & \colhead{Flux} 
 \\
\colhead{}  & \colhead{Wavelength)}  & \colhead{exposure} & 
\colhead{}  & \colhead{\hfill(units 10$^{-16}$} & \colhead{erg cm$^{-2}$ s$^{-1}$ \AA$^{-1}$)} 
 \\
\colhead{}  & \colhead{(\AA)}  & \colhead{(s)} & 
\colhead{}  
 \\

}
\startdata
V & 5430 & 80     & 1.05 $\pm$ 0.18 & 2.64 & $2.76 \pm 0.48$  \\
B & 4340 & 80     & 1.31 $\pm$ 0.22 & 1.33 & $1.75 \pm 0.29$  \\
U & 3440 & 90        & $<$0.36  & 1.54 & $<0.54 $ \\
UVW1 & 2600 & 90     & $<$0.15  & 3.37 & $<0.46 $ \\
UVM2 & 2200 & 90     & $<$0.09  & 7.60 & $<0.72 $ \\
UVW2 & 1930 & 80     & $<$0.09  & 4.94 & $<0.50 $ \\

\enddata
\tablecomments{upper limits are quoted at 3$\sigma$. The zeropoint is the flux corresponding to 1 c/s.}
\end{deluxetable}

Examination of the UVOT finding chart exposure reveals a new source within the XRT positional error circle of the burst 
(Krimm et al. 2005a)  at RA = 10h 16m 47.76(3)s Dec = +43d 32m 54.9(5)s J2000. The source fades with time, demonstrating that it is 
the afterglow of the burst. It has a mean V magnitude 
of 17.5 during the exposure, which is centered 140s after the burst trigger (278s after burst onset). The position of the afterglow is consistent with that 
reported by Rykoff et al. (2005). The afterglow is also seen in the UVOT B filter, in which the first 
observation is a ten second exposure centered 243s after the trigger. It is not present in observations made with any of the
shorter wavelength UVOT filters taken around the same time. 

The brightness of the afterglow was
measured by extracting the counts from a circular aperture of 6 arc seconds radius centered on the source position. 
The background was 
measured from an annular region surrounding the source that had an inner radius of 
6 arc sec and an outer radius of 30 arc seconds. Pixels in the background region
that were significantly above the mean (due for example to the presence of another star) were rejected. The background
count rate varies with Earth aspect angle but is typically 0.01 c/s/arcsec$^2$ in V and 0.02 c/s/arcsec$^2$ in B.
In Figure~2 we show the earliest UVOT data, plotted as a detector count rate on a linear scale in 10s bins. 
These data, taken in event mode, extend to 1700s after 
the trigger. We include both the V-band and B-band data. By coincidence the count rate in the two filters is very similar, 
the greater
transmission of the instrument in the B band being compensated for by the fact that source is brighter in V. If we compare the
count rate in the interval 250-950s after the trigger, when the instrument was cycling often through the two filters, 
we get a mean count rate of 
1.05$\pm$0.18 c/s in V and 1.31$\pm$0.22 c/s in B. These are consistent within the 1 sigma error bounds, but we have 
multiplied the B-band data in Figure~2 by the mean factor of 0.8 for consistency. 
A power-law fit to the combined V-band and B-band data, shown as a solid line in the 
figure, has a slope of $\alpha = -0.57 \pm 0.07$. We have included the UVOT datapoints beyond 1700s in this fit. The origin of the
power law in this treatment, $T_0$ is taken to be the time of the BAT trigger. However, as noted above, the burst actually began 
about 138s before the trigger, so the appropriate value of $T_0$ is debateable. We have fit the data allowing $T_0$ to be a 
free parameter, which yields $T_0=-54 \pm 73$s, and a slightly steeper slope of $\alpha = -0.63 \pm 0.08$. This fit is shown as a 
dashed line in Figure~2, but is statistically indistinguishable from the $T_0 = 0$ case.
The reduced $\chi^2$ value for the power law fit to the time evolution data in 10s bins is 0.75 per d.o.f., which is 
statistically acceptable. 

Figure~3 shows the B and V magnitude of the afterglow as a function of time since the trigger. 
The data, listed in Table~1, 
have been binned
in time with respect to Figure~2 to increase signal to noise.
The instrumental count rates
are converted to magnitudes and fluxes using preliminary on-orbit calibration data, which might be subject to slight revision
as calibration knowledge improves. The zeropoints used are listed in Figure~2. 
The magnitudes quoted here differ significantly, particularly in the V-band, from those reported 
by Boyd et al. (2005), which were derived before on-orbit calibration data were available and were therefore
based only on ground calibration estimates of the zeropoints. The UVOT magnitude system is defined such that the magnitude of 
Vega would be zero in each filter.

The $\alpha = -0.57$ power law derived above is compared to both the V-band and B-band data in Figure~3. 
Both wavebands are consistent with the same overall slope. Specifically, the UVOT data indicate  a B-V color of $\sim$1 with no evidence of systematic evolution with time.  
We also show ground-based telescope data reported by Yoshioka et al. (2005), Quimby et al (2005), Kiziloglu et al. (2005), Misra et al. (2005), 
Sharapov et al. (2005a,b), Greco et al. (2005) and Wo\'zniak et al (2005), most of which report R-band or broadband 
measurements referenced to
the R-band magnitudes of field stars. Comparison of these data with
those from UVOT suggest V-R$\sim$0. However we caution that the transformation of broadband measurements into standard color systems
by reference to field stars is subject to systematic uncertainty, particularly, as in this case, when the intrinsic spectrum of the
object of interest is likely to be different from that of the average field star. 

It is interesting that the decay slope derived from the UVOT data extrapolates well to the early measurement
of Quimby et al. (2005) and Wo\'zniak et al (2005); the former authors find R=16.16$\pm$0.17 27s after the Swift trigger, 
while the latter quote R=16.323$\pm$0.046 40s after the trigger. Wo\'zniak et al (2005) find that there are systematic 
deviations from a simple power law during the early decline phase. Though of lower statistical quality, the UVOT data 
are very consistent with these data in suggesting an excess above the power law in the interval 200-800s after the trigger.
Wo\'zniak et al  fit a broken power law decay to their data (with T$_{\rm break}\sim 400-500$s). However the later time UVOT 
and ground based measurements suggest that these deviations can be characterised as fluctuations about a 
mean $\alpha \sim -0.5$ slope. 

At late times, analysis of the Swift XRT data by Cusumano et al. (2005) indicate that there is a break in the X-ray light curve
of GRB~050319 at about $2.6 \times 10^4$ sec, from a slope consistent with that of the optical data, to one with $\alpha = -1.14$. 
Such a break is indicated in Figure~3 for illustration. While the UVOT data do not constrain the presence of such a break, the
late ground based optical data  seem to favour an extension of the $\alpha = -0.5$ slope.

\subsection{Spectrum}
To examine the spectrum of GRB050319, we use data from the interval between 240s and 930s after the trigger, when the 
UVOT was cycling though each of its six color filters with exposure times of 10s per filter, ensuring that the data from each filter
sample the same portion of the decay light curve. The filter wheel was rotated 8.5 times during this interval, yielding 8 observations
in the V, B and UVW2 filters, and nine in each of U, UVW1 and UVM2. The mean count rate and flux, or three sigma limit on these
quantities, is listed in Table~2. The spectral flux distribution is shown in Figure~4. 
Reddenning due to the Galaxy in the direction
of the burst is very low. At A$_{\rm V}=0.034$ (Schlegel, Finkbeiner \& Davis 1998), no correction is applied in the plot.   
We also show X-ray data from the Swift XRT averaged
over exactly the same time interval as the UVOT data. The count data derived from the XRT during this time period are fitted
to a power law spectrum absorbed by a fixed Galactic absorbing column of 
N$_{\rm H}$=$1.17 \times 10^{20}$ cm$^{-2}$. The best fit spectrum
has a slope ($\beta$) of $-0.8 \pm 0.1$. This fit is used to convert the count rate data to incident flux, and these data are then
plotted in Figure~4, corrected for the Galactic absorption, to compare with the UVOT measurements. 

A simple power law extrapolation of the XRT X-ray spectrum predicts the V-band flux well.  
However, the UVOT B-band measurement at 435nm
and the upper limits to the flux in the U, UVW1, UVM2 and UVW2 filters, which cover the wavelength range 180-390 nm, 
lie significantly
below any reasonable distribution joining the V-band and X-ray data. 
The B-band flux is lower than the power law joining X-ray and V-band by a factor of 
$\sim$2. This suggests that the optical spectrum is cut-off in
the B band. If this is solely due to the Lyman edge at 91.2 nm in the rest frame, the implied redshift of the burst is $\sim$3.8. 
This is effectively
an upper limit to the burst redshift since there may be additional sources of opacity in the spectrum longward of the 
Lyman edge.  
For example 
Lyman $\alpha$ absorption in the host galaxy or the Lyman forest could contribute to the flux deficit. 
Fynbo et al. (2005) report an absorption line system at a redshift of 3.24 in the spectrum of the afterglow. This 
would imply a Lyman edge at 387nm, shortward of 97\% of the response of
the UVOT B-band filter. The absorption line system is not necessarily due to the host galaxy of the burst, so conservatively we could 
bracket the burst redshift between 3.24 and $\sim$3.8.
However the UVOT data would be consistent with a host galaxy redshift of 3.24
provided there is indeed significant additional absorption, due to Lyman $\alpha$ for example, in the B band. Based on the
IGM transmission model of Madau (1995) the transmission through the B filter is consistent 
with z=3.24 up to 3.5 at the $\pm$1 sigma level, and with z$<$3.8 at the 3 sigma level.

\section{Discussion}

%

The UVOT instrument detected and followed the optical afterglow of GRB050319 from when it was acquired by the Swift 
narrow-field instruments, about 90s after the burst trigger, until it faded below the UVOT sensitivity threshold, about
50,000s later. The afterglow was detected in the V and B bands, but not at shorter wavelengths. Interpreting this 
spectral cut-off as being due to the Lyman edge of hydrogen suggests an upper limit to the burst redshift of $\sim 3.8$.
This is consistent with the reported absorption line system at z=3.24 (Fynbo et al. 2005), which is likely to be
the host galaxy of the burst.
There is no evidence for significant evolution of the B-V color as a function of the burst decay
(Figure 3).

The multiwavelength spectrum of the afterglow, derived from the interval between 240s and 940s after the trigger, 
suggests that the
observed V-band flux (which is at a wavelength of $\sim$ 1300\AA\ in the rest frame of the burst) 
is an extension of the $\beta = -0.8 \pm 0.1$ X-ray spectral distribution. 
Indeed the 
combined X-ray and V-band data are perfectly consistent with this slope. The interval used to compile this spectrum
begins just after the first break in the X-ray time evolution curve (see below) as reported by Cusumano et al. (2005). The 
X-ray spectrum before this break is significantly softer ($\beta= -1.6$) than after ($\beta= -0.7$) and the interval used
for the X-ray spectrum in Figure~4 may be slightly contaminated by the remnants of the soft component. 
However Cusumano et al. derive a weighted average spectral slope for all the X-ray data after this break 
of $\beta = -0.73 \pm 0.05$, which is consistent with our X-ray -- V-band slope within errors, suggesting 
any such contamination is minor.  
The multiwavelength spectrum of GRB~050319 is in accord with expectations from the relativistic fireball 
model (e.g. Granot \& Sari 2002) and suggests that a single optically thin synchrotron afterglow spectrum 
extends from the X-ray to the (rest frame) UV band. This idea 
is supported by the fact that the mean optical decay slope is indistinguishable from the 
decay slope of the X-ray flux
in the interval between about 230s and $2.7 \times 10^{4}$s after the trigger, which is $\alpha = -0.51 \pm 0.03$ 
(Cusumano et al. 2005).

The optical flux decays with a mean power law slope of $\alpha = -0.57 \pm 0.07$, but there is 
good evidence for fluctuations from this
steady decline, from the UVOT and particularly from the well-sampled early phase coverage of
Wo\'zniak et al. (2005). Such variability might be due to fluctuations in the
density of the circum-burst medium, or due to continued injection of relativistic material into the 
fireball (refreshed shock). The shallow overall decay slope, combined with the observed 
spectral slope would require an unusually flat 
electron energy
spectral index in the context of the simple fireball model (e.g. Granot \& Sari 2002). 
However the idea 
that the central engine is continuously injecting energy into 
the fireball may be a more natural explanation of the shallow decay.  If we parameterize the injection as 
producing a luminosity $L \propto t^{-q}$, this
continuous injection would influence the dynamics of the fireball, and
hence the lightcurves, as long as $q<1$ (Zhang \& M\'esz\'aros 2001). Assuming that the
central engine stops injecting energy after $2.7 \times 10^4$s, we 
get a consistent solution for the UVOT and XRT lightcurve with $q$ in the range
$0.5-0.6$ and $p \sim 2.8$. Both the optical and the X-ray bands are
between the injection frequency, $\nu_m$ and the cooling frequency, $\nu_c$ throughout. This is satisfied within a wide
parameter regime.

Interestingly, prior to T+230s the X-ray flux decays much more steeply, with a power law slope of 
$\alpha = -2.9 \pm 0.3$ (Cusumano et al. 2005). Thus at T+90s, when the burst was first acquired by the 
Swift narrow field instruments,
the X-ray flux was more than 20 times brighter than it was at T+230s. There is no evidence for a similar rapidly decaying
component in the flux measured with the UVOT, which would have needed to be at about 14$^{th}$ magnitude 
rather than the observed 17$^{th}$ magnitude at T+100s to match the 
XRT data. Furthermore, the ROTSE  and Raptor data taken as early as 27s after the BAT trigger 
(Quimby et al. 2005; Wo\'zniak et al. 2005)  
are also consistent with an
extrapolation of the $\alpha = -0.57$ UVOT decay profile to earlier times with no rapidly decaying component.
This is firm evidence that the early X-ray
afterglow lightcurve is composed of two distinct components. The steep
decay is likely to be the tail emission from the internal shocks (mainly due to the
so-called curvature effect; Kumar \& Panaitescu 2000), while the
shallow broadband decay component comes from
the external shock afterglow. 

Early rapid decay phases have now been seen in a number of
bursts with the XRT (cf. Cusumano et al. 2005).
Optical observations have been made while the Gamma-ray burst is ongoing in two bursts thus far, 
GRB990123 (Akerlof et al. 1999) and 
GRB041219 (Vestrand et al. 2005). In the case of GRB990123 the optical emission was not correlated in 
time with the gamma-rays, peaking 
instead approximately 20s after the gamma-ray emission. A possible explanation for the optical 
light in this case is that it arose in 
the reverse shock driven in to the ejecta by interaction with the surrounding medium 
(M\'esz\'aros \& Rees 1997; Sari \& Piran 1999; Zhang, Kobayashi
\& M\'esz\'aros 2003). 
However, in the case of
GRB041219, the optical emission was well corellated with the gamma-rays, suggesting that it arose due to internal shocks in the
ultra-relativistic ejecta itself (Vestrand et al. 2005). 
If the rapidly decaying XRT flux in GRB050319 is related 
to the prompt emission, then neither mechanism resulted in a detectable optical flash 
in GRB050319. This might be due to various mechanisms that
suppress the reverse shock optical flash (e.g. Kobayashi 2000; Zhang
\& Kobayashi 2005), and because of the strong synchrotron
self-absorption effect during the prompt emission phase which
suppresses the internal shock optical flash (Fan et al. 2005).

\acknowledgments
We thank John Cannizzo for supplying the BAT data that was used in Figure~1.
The Swift programme is supported by NASA, PPARC and ASI.

\end{document}